\documentclass[iop]{emulateapj}

\newcommand{\cf}{{\ifmmode{C_{\rm f}}\else{$C_{\rm f}$}\fi}}
\newcommand{\zem}{{\ifmmode{z_{\rm em}}\else{$z_{\rm em}$}\fi}}
\newcommand{\zabs}{{\ifmmode{z_{\rm abs}}\else{$z_{\rm abs}$}\fi}}
\newcommand{\zs}{{\ifmmode{z_{\rm s}}\else{$z_{\rm s}$}\fi}}
\newcommand{\zl}{{\ifmmode{z_{\rm l}}\else{$z_{\rm l}$}\fi}}
\newcommand{\kms}{{\ifmmode{{\rm km~s}^{-1}}\else{km~s$^{-1}$}\fi}}
\newcommand{\vej}{{\ifmmode{v_{\rm ej}}\else{$v_{\rm ej}$}\fi}}
\newcommand{\vrot}{{\ifmmode{v_{\rm rot}}\else{$v_{\rm rot}$}\fi}}
\newcommand{\cm}{{\ifmmode{{\rm cm}^{-1}}\else{cm$^{-1}$}\fi}}
\newcommand{\cmm}{{\ifmmode{{\rm cm}^{-2}}\else{cm$^{-2}$}\fi}}
\newcommand{\cmmm}{{\ifmmode{{\rm cm}^{-3}}\else{cm$^{-3}$}\fi}}
\newcommand{\lya}{Ly$\alpha$}

\newcounter{species} 
\def\ion#1#2{\setcounter{species}{#2}#1$\;${\scriptsize\Roman{species}}\relax}

\shorttitle{Spectroscopic Observation of SDSS~J1001+5027}
\shortauthors{Misawa et al.}

\begin{document}

\title{Spectroscopic Observations of the Outflowing Wind in the Lensed
  Quasar SDSS~J1001+5027\altaffilmark{1}}

\footnotetext[1]{Based on data collected at Subaru Telescope, which is
  operated by the National Astronomical Observatory of Japan.}

\author{Toru Misawa$^2$, 
        Naohisa Inada$^3$, 
        Masamune Oguri$^{4,5,6}$,
        Jane C. Charlton$^7$, 
        Michael Eracleous$^{7,8}$, 
        Suzuka Koyamada$^9$, 
    and Daisuke Itoh$^9$
}

\affil{
$^2$School of General Education, Shinshu University, 3-1-1 Asahi,
  Matsumoto, Nagano 390-8621, Japan; misawatr@shinshu-u.ac.jp \\
$^3$Department of Physics, National Institute of Technology, Nara
  College, Yamatokohriyama, Nara 639-1080, Japan \\
$^4$Research Center for the Early Universe, University of Tokyo, 7-3-1
  Hongo, Bunkyo-ku, Tokyo 113-0033, Japan \\
$^5$Department of Physics, University of Tokyo, 7-3-1 Hongo,
  Bunkyo-ku, Tokyo 113-0033, Japan \\
$^6$Kavli Institute for the Physics and Mathematics of the Universe
  (Kavli IPMU, WPI), University of Tokyo, Chiba 277-8583, Japan \\
$^7$Department of Astronomy \& Astrophysics, The Pennsylvania State
  University, 525 Davey Lab, University Park, PA 16802 \\
$^8$Institute for Gravitation and the Cosmos, The Pennsylvania State
  University, 525 Davey Lab, University Park, PA 16802 \\
$^9$Department of Physics, Faculty of Science, Shinshu University,
  3-1-1 Asahi, Matsumoto, Nagano 390-8621, Japan \\
}

\begin{abstract}
We performed spectroscopic observations of the small-separation lensed
quasar SDSS~J1001+5027, whose images have an angular separation
$\theta = 2.^{\!\!\prime\prime}86$, and placed constraints on the
physical properties of gas clouds in the vicinity of the quasar (i.e.,
in the outflowing wind launched from the accretion disk).  The two
cylinders of sight to the two lensed images go through the same region
of the outflowing wind and they become fully separated with no overlap
at a very large distance from the source ($\sim$330~pc).  We
discovered a clear difference in the profile of the \ion{C}{4} broad
absorption line (BAL) detected in the two lensed images in two
observing epochs.  Because the kinematic components in the BAL profile
do not vary in concert, the observed variations cannot be reproduced
by a simple change of ionization state.  If the variability is due to
gas motion around the background source (i.e., the continuum source),
the corresponding rotational velocity is $v_{\rm rot}\geq
18,000$~\kms, and their distance from the source is $r\leq 0.06$~pc
assuming Keplerian motion.  Among three \ion{Mg}{2} and three
\ion{C}{4} NAL systems that we detected in the spectra, only the
\ion{Mg}{2} system at \zabs\ = 0.8716 shows a hint of variability in
its \ion{Mg}{1} profile on a rest-frame time scale of $\Delta t_{\rm
  rest}$ $\leq$ 191~days and an obvious velocity shear between the
sightlines whose physical separation is $\sim$ 7~kpc. We interpret
this as the result of motion of a cosmologically intervening absorber,
perhaps located in a foreground galaxy.
\end{abstract}


\section{Introduction}
AGN outflows, which could be powered by various mechanisms (e.g.,
radiation or magnetic pressure, and magnetocentrifugal forces), are a
very important process in the evolution of quasars as well as the
evolution of their host galaxies because 1) they may facilitate the
extraction of angular momentum from accreting gas, allowing gas
accretion to proceed \citep[e.g.,][]{bla82,emm92,kon94,eve05}, 2) they
provide energy and momentum feedback to the interstellar medium (ISM)
of host galaxy and to the intergalactic medium (IGM), and inhibit star
formation activity \citep[e.g.,][]{spr05}, and 3) they may enrich the
IGM with heavy elements \citep[e.g.,][]{ham97b, scan04, gab06,
  shenmadau12}.

The outflowing matter from quasars is detected via absorption features
in their spectra, which appear in addition to the plethora of
absorption features produced by intervening objects (e.g., foreground
galaxies, the IGM, and clouds in the Milky Way).  The former, usually
called {\it intrinsic} absorption lines, are observed in the spectra
of about half of all quasars \citep[e.g.,][]{ves03, wis04, mis07a,
  nes08}, while the latter, usually called {\it intervening}
absorption lines, are detected in virtually all quasar spectra.

Intrinsic absorption lines are a powerful tool for probing the
outflowing winds of quasars, which are difficult to observe directly
\citep[e.g.,][]{wey91, ham97a, ara99, sri02, moe09, cap11, fil13}.
However, a drawback of most studies of intrinsic quasar absorption
lines is that they probe {\it only} a single sight line (i.e., one
dimension) toward the nucleus of each quasar, while the physical
conditions of the outflow are likely to depend on the polar angle
relative to the axis of the quasar \citep[e.g.,][]{gan01,elv00}.
Therefore, the structure of the outflowing winds (e.g., absorber's
size, kinematic motions, and volume density) is still largely unknown.

Multiple images of quasars produced by gravitational lensing provide a
unique way for studying multiple sightlines through the same outflow
\citep[e.g.,][]{tur01}.  Lensed quasars with large image separation
angles have a very good chance of revealing structural differences in
the outflowing winds in the vicinity of the quasars themselves.  In
this sense, the following three lensed quasars are the most promising
targets because they are lensed by a cluster of galaxies rather than
by a single massive galaxy: SDSS~J1004+4112 with separation angle of
14$^{\prime\prime}\!\!$.6 \citep{ina03}, SDSS~J1029+2623 with
22$^{\prime\prime}\!\!$.5 \citep{ina06,ogu08}, and SDSS~J2222+2745
with 15$^{\prime\prime}\!\!$.1 \citep{dah13}.

Among these, SDSS~J1029+2623 at $z_{em}=2.197$, the current
record-holding lensed quasar whose images have the largest known
angular separation, has proven to be an excellent target for the study
of the structure of its outflow because there are clear absorption
features detected at the blue wings of the \ion{C}{4}, \ion{N}{5}, and
Ly$\alpha$ emission lines in the spectra of the lensed images
\citep{ina06,ogu08}.  \citet{mis13} obtained a high-resolution
spectrum using the Subaru telescope and the High-Dispersion
Spectrograph (HDS), and discovered several clear indications that the
absorption lines indeed arise in the outflowing wind; partial
coverage\footnote[10]{Absorbers do not cover the background flux
  source completely along our sightline.}  and
line-locking\footnote[11]{The red component of a doublet (such as
  \ion{C}{4}, \ion{Si}{4}, and \ion{N}{5}) is aligned with a blue
  component of the following doublet, which is a signature of a
  radiatively driven wind and only detectable if our sightline is
  almost parallel to the wind streamlines
  \citep{ben05,bow14}}. \citet{mis13} also discovered a clear
difference between the broad absorption profiles in the spectra of the
two lensed images.

We observed SDSS~J1029+2623 again using Subaru/HDS \citep{mis14b} and
the Very Large Telescope (VLT) with the Ultraviolet and Visual Echelle
Spectrograph (UVES) \citep{mis16} about four years ($\sim$1460~days)
after the first observation \citep[which is longer than the
  $\sim$774~day time-delay between the lensed images;][]{foh13}.  We
confirmed that the difference in absorption profile persists at the
3$\sigma$ level in the new spectra, which suggests that our sightlines
to the lensed images are indeed passing through different regions of
the outflowing wind at different angles (i.e., they are not a result
of time variability due to the time delay between the images)
\citep{mis14b}.  The size of an absorbing gas cloud, $d$, at a
distance of $r$ from the source should be smaller than the physical
distance between the sightlines of the two lensed images, i.e., $d$
$\leq$ $r\,\theta^{\prime}$, to avoid covering the cylinders of both
sightlines. In the above expression $\theta^{\prime}= (D_{\rm
  ol}/D_{\rm sl})\,\theta$ is the separation angle seen from the
source with $\theta$ the observed angular separation between the
images and $D_{\rm ol}$ and $D_{\rm sl}$ the angular diameter
distances from the observer to the lens and from the source to the
lens, respectively. This result places a direct constraint on the
absorber's size in the direction perpendicular to the line of sight
for the first time.

In this paper, we carry out the same type of observations for the
quasar, SDSS~J1001+5027, which is lensed not by a cluster of galaxies
but by a single massive galaxy.  Because the lensed images of this
quasar have an angular separation of $\sim$3$^{\prime\prime}$ (about
ten times smaller than that of SDSS~J1029+2623), the cylinders of
sight to the two lensed images overlap almost completely at a small
distance from the source. Thus we cannot observe the structure of the
outflow from two different directions. Instead, we can effectively
monitor the outflow at two epochs by virtue of the time-delay between
the lensed images.  Among $\sim$150 lensed quasars that have been
discovered so far in the CfA-Arizona Space Telescope LEns
(CASTLE)\footnote[12]{https://www.cfa.harvard.edu/castles/} and SDSS
Quasar Lens Search
(SQLS)\footnote[13]{http://www-utap.phys.s.u-tokyo.ac.jp/~sdss/sqls/}
surveys, only SDSS~J1001+5027 satisfies all of the following criteria:
1) strong absorption troughs exist in the blue wings of broad emission
lines (i.e., a promising candidate for blueshifted lines from an
outflowing wind), 2) bright enough ($i$ $<$ 18), 3) the angular
separation of the lensed images is at least three times the typical
seeing size at Mauna Kea ($\theta$ $\ga$ 2.$^{\!\!\prime\prime}$0), 4)
good target visibility from the Subaru telescope, and 5) good
sensitivity in the optical band for the detection of \ion{C}{4}
absorption lines at $\lambda_{\rm obs}$ $\sim$ 4000 --
7000~\AA\ (i.e., $z_{em}$ $\sim$ 1.6 -- 3.5).  Our goals are ($i$) to
place new limits on the physical properties of the BAL absorbing
clouds and their distance from the source and ($ii$) to identify
narrow absorption lines (NALs) that originate in the outflowing wind
based on identical line profiles measured in spectra of comparably
high signal-to-noise ratio (hereafter S/N) from the two different
sightlines.

The paper is organized as follows.  We describe the detailed
properties of our target SDSS~J1001+5027 in \S2 and the observations
and data reduction in \S3.  The results and discussion are presented
in \S4 and \S5. Finally, we summarize our conclusions in \S6.  We use
a cosmology with $H_{0}$=70 \kms~Mpc$^{-1}$, $\Omega_{m}$=0.3, and
$\Omega_{\Lambda}$=0.7 throughout the paper.

\section{The Lensed Quasar SDSS~J1001+5027}
The quasar SDSS~J1001+5027 at $\zem=1.838$ was first reported in
\citet{ogu05} as a lensed quasar with two images, A ($i$ = 17.36) and
B ($i$ = 17.71), that had been originally selected as a lens candidate
from the Sloan Digital Sky Survey (SDSS) \citep{yor00}.  The redshift
of the main lensing galaxy is spectroscopically determined as \zl\ =
0.415 \citep{ina12}.  The image separation
($\theta=2^{\prime\prime}\!\!.86$) is slightly larger than the typical
value for lensed quasars by a single massive galaxy ($\theta$
$\sim$1--2$^{\prime\prime}$).  \citet{ogu05} discovered a possible
second lensing galaxy and an enhancement of galaxy number density on a
larger scale, within 60$^{\prime\prime}$ of the quasar images.  These
additional galaxies could contribute to the large image separation.

Given the redshifts of the lensing galaxy and the source (i.e., the
quasar), the separation angle of the light rays that form the images~A
and B, {\it as seen from the source}, is evaluated as
$\theta^{\prime}= 1.\!\!^{\prime\prime}37$, which is almost half the
separation angle seen from our position.  The time-delay between the
images has also been measured.  \citet{rat13} monitored the optical
$R$-band light curves of images~A and B with measurements at 443
independent epochs for more than six years, and concluded that image~A
leads B by 119.3$\pm$3.3~days after combining results from five
different methods.\footnote[14]{\citet{agh17} also evaluated the
  time-delay as 117$^{+7.1}_{-3.7}$~days based on their own algorithm,
  which is in good agreement with the result by \citet{rat13}.}
\citet{rus16} also obtained several possible mass models for the
lensing object and determined a total magnification factor
$\mu=3.51$\footnote[15]{This is a weighted average of magnification
  factors for their 13 mass models.} of the quasar images. Because the
flux ratio of images~A and B is 1:0.77 \citep{ogu05}, we will use
$\mu_{\rm A}$ = 1.99 and $\mu_{\rm B}$ = 1.52 as magnification factors
for images~A and B respectively throughout the paper.

Here, we estimate some of the physical parameters of the quasar that
will be useful in our later discussion on our results.  We first
estimate the sizes of the broad emission line region (BELR), $R_{\rm
  BELR}$, and the continuum source, $R_{\rm cont}$, following
\citet{mis13}.  The former is calculated using the empirical relation
between $R_{\rm BELR}$ and quasar luminosity \citep{mcl04}.  The
monochromatic luminosity of image~A at $\lambda_r$ = 3000\AA\ in the
quasar rest-frame is measured by \citet{she11} as $\log \lambda
L_{3000}$ = 46.52$\pm$0.01 erg~s$^{-1}$. After correcting for
magnification we estimate $R_{\rm BELR}$ $\sim$0.4$^{+0.7}_{-0.3}$~pc,
where the main source of uncertainty is the scatter in the empirical
relation in \citet{mcl04}.  As to the size of the continuum source, we
take five times the Schwarzschild radius ($5R_S$ = 10$GM_{BH}/c^2$)
following \citet{mis05}, where $M_{BH}$ is the mass of the central
black hole of the quasar. Recent microlensing studies suggest that the
size of the continuum source is larger than the estimate based on the
Shakura-Sunyaev disk model \citep{sha73} by a factor of four
\citep[e.g.][see also the discussion in \citealt{rod13}]{mor10}.  In
this paper, however, we use the original value without correcting it
so as to compare the results of our current target SDSS~J1001+5027 to
those of our previous target SDSS~J1029+2623
\citep{mis13,mis14b,mis16}.  The virial mass of the central black hole
is already calculated in \citet{she11}. After correcting for
magnification, we obtain a black hole mass of $\log
(M_{BH}/M_{\odot})$ = 9.66$\pm$0.06 and a continuum source size of
$R_{\rm cont} = (2.2\pm 0.3)\times10^{-3}\;$pc, about 170 times
smaller than $R_{\rm BELR}$.  The Eddington ratio of the quasar
corresponding to the above values is $L/L_E$ $\sim$ 0.15, which is a
value of typical bright quasars, $L/L_E$ $\sim$ 0.07 -- 1.6
\citep{net07}.

Based on the separation angle seen from the source, $\theta^{\prime}$,
and the size of sources, $R_{\rm BELR}$ and $R_{\rm cont}$, we are
also able to calculate the boundary distance, $r_{\rm b}$, i.e., the
distance from the source where the two cylinders of sight for an
extended object corresponding to the two images become fully separated
with no overlap as introduced in \citet{mis13}.  The boundary distance
for SDSS~J1001+5027 is $r_b = 330^{+50}_{-40}$~pc if only the
continuum source is the flux source or $60^{+100}_{-40}$~kpc if the
BELR is also part of the background source.  Both of these are
considerably larger than those evaluated for the large-separation
lensed quasar SDSS~J1029+2623 \citep[$\sim$3.5~pc and $\sim$1.2~kpc,
  respectively; see][]{mis16}.

% =================================================================
% Figure 1
% =================================================================
\begin{figure*}
 \begin{center}
  \includegraphics[width=15cm,angle=0]{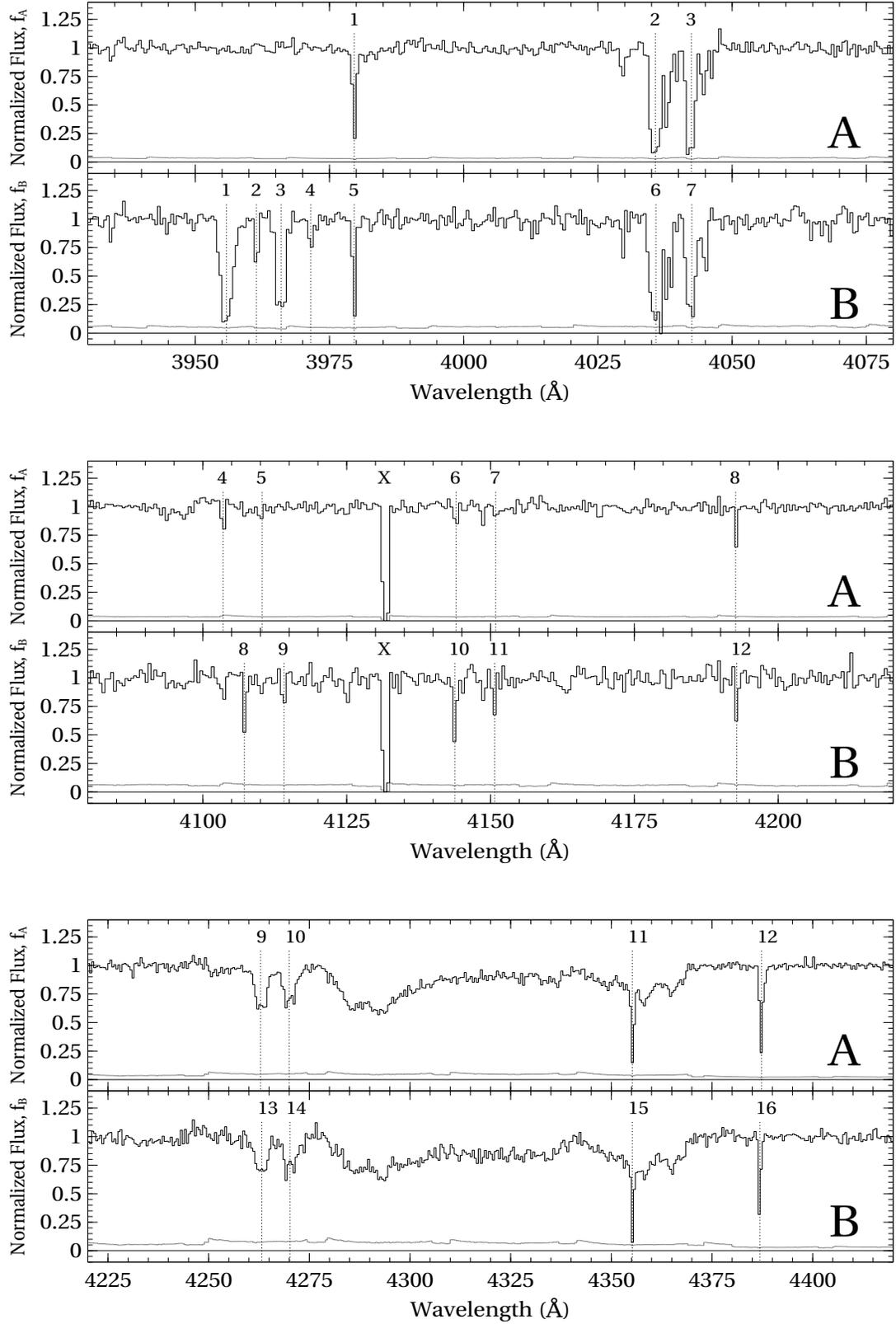}
 \end{center}
\caption{Normalized spectra and their 1$\sigma$ errors (lower traces)
  for images~A and B of SDSS~J1001+5027 in epoch E1, taken with
  Subaru/HDS, after sampling every 0.5\AA\ for display purposes only.
  Identification numbers from Table~\ref{tab:abslines} are shown above
  the corresponding absorption lines.  Crosses denote data defects
  such as bad columns and bad pixels.\label{fig:spec}}
\end{figure*}

\begin{figure*}
\addtocounter{figure}{-1}
 \begin{center}
  \includegraphics[width=15cm,angle=0]{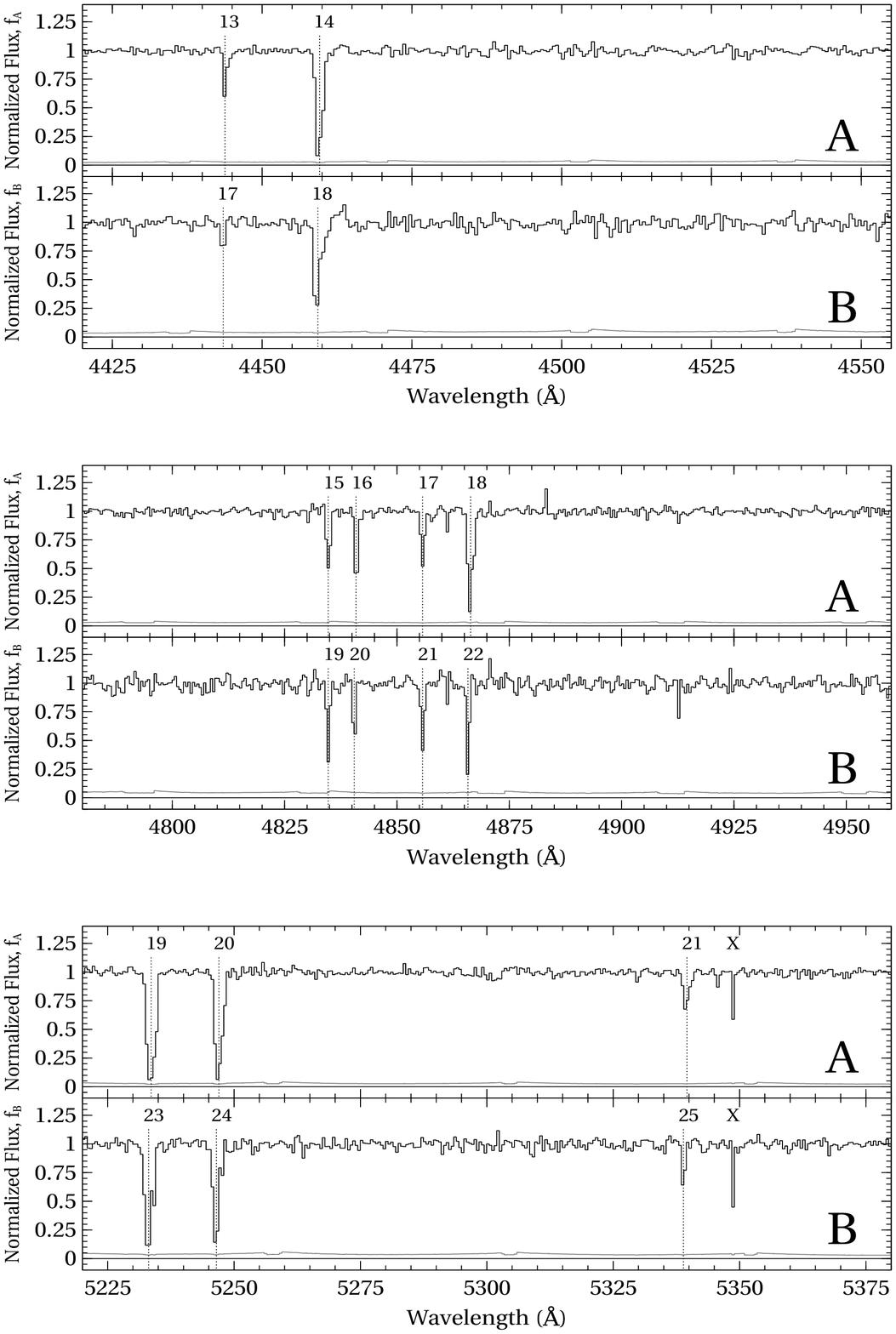}
 \end{center}
 \caption{Continued.}
\end{figure*}
% =================================================================

\section{Observations and Data Reduction}
We conducted high-resolution spectroscopic observations of the two
images of the lensed quasar SDSS~J1001+5027 using Subaru/HDS on 2016
January 27 and 2017 January 19 (epochs E1 and E2, hereafter) whose
time separation ($\sim$358~days) is larger than the time delay between
the images ($\sim$119~days) in the observed frame.  We used a slit
width of $0.\!\!^{\prime\prime}8$, which yielded a resolving power of
$R$ $\sim$ 45,000.  The wavelength coverage is 3820--4600 \AA\ on the
blue CCD and 4680--5490~\AA\ on the red CCD, which includes the
\ion{Si}{4}, \ion{C}{4}, and \ion{C}{3}] emission lines at the
  redshift of this quasar.  Because the spectra were oversampled
  (i.e., there were 8 pixels per resolution element), we binned every
  4 pixels in both spatial and dispersion directions (i.e.,
  $\sim$0.05\AA\ per bin) to increase the S/N.  The sky conditions
  were good. The total integration time was 12,000~s for each image in
  each epoch, except for image~B in epoch E1 whose integration time
  was slightly longer, $\sim$14400~s.

We reduced the data in a standard manner with the IRAF
software\footnote[16]{IRAF is distributed by the National Optical
  Astronomy Observatories, which are operated by the Association of
  Universities for Research in Astronomy, Inc., under cooperative
  agreement with the National Science Foundation.}.  Wavelength
calibration was performed with the help of a Th-Ar lamp spectrum.  We
carried out the flux calibration of the spectra using the spectrum of
the spectrophotometric standard star Feige~34. The final S/N is about
9--16~bin$^{-1}$ at 4500\AA.  The log of the observations is presented
in Table~\ref{tab:obslog}.

% =================================================================
% Table 1
% =================================================================
\begin{deluxetable*}{cccccccc}
\tablecaption{Log of Observations \label{tab:obslog}}
\tablewidth{0pt}
\tablehead{
\colhead{Target}         &
\colhead{RA}             &
\colhead{Dec}            &
\colhead{$m_{\rm i}$}      &
\colhead{Obs Date}       &
\colhead{$R$}            &
\colhead{$T_{\rm exp}$}    &
\colhead{S/N$^a$}        \\
\colhead{}               &
\colhead{(hh mm ss)}     &
\colhead{(dd mm ss)}     &
\colhead{(mag)}          &
\colhead{}               &
\colhead{}               &
\colhead{(sec)}          & 
\colhead{(bin$^{-1}$)}    
}
\startdata
SDSS~J1001+5027~A & 10 01 28.61 & +50 27 56.9 & 17.36 & 2016 Jan 27 & 45000 & 12000 & 16 \\
                  &             &             &       & 2017 Jan 19 & 45000 & 14400 & 15 \\
\hline
SDSS~J1001+5027~B & 10 01 28.35 & +50 27 58.5 & 17.71 & 2016 Jan 27 & 45000 & 12000 & 11 \\
                  &             &             &       & 2017 Jan 19 & 45000 & 12000 &  9 
\enddata
\tablenotetext{a}{Signal to noise ratio at $\lambda_{\rm obs}$ $\sim$
  4500\AA.}
\end{deluxetable*}
% =================================================================

\section{Results}
\begin{figure}[h]
\begin{center}
  \includegraphics[width=8cm,angle=0]{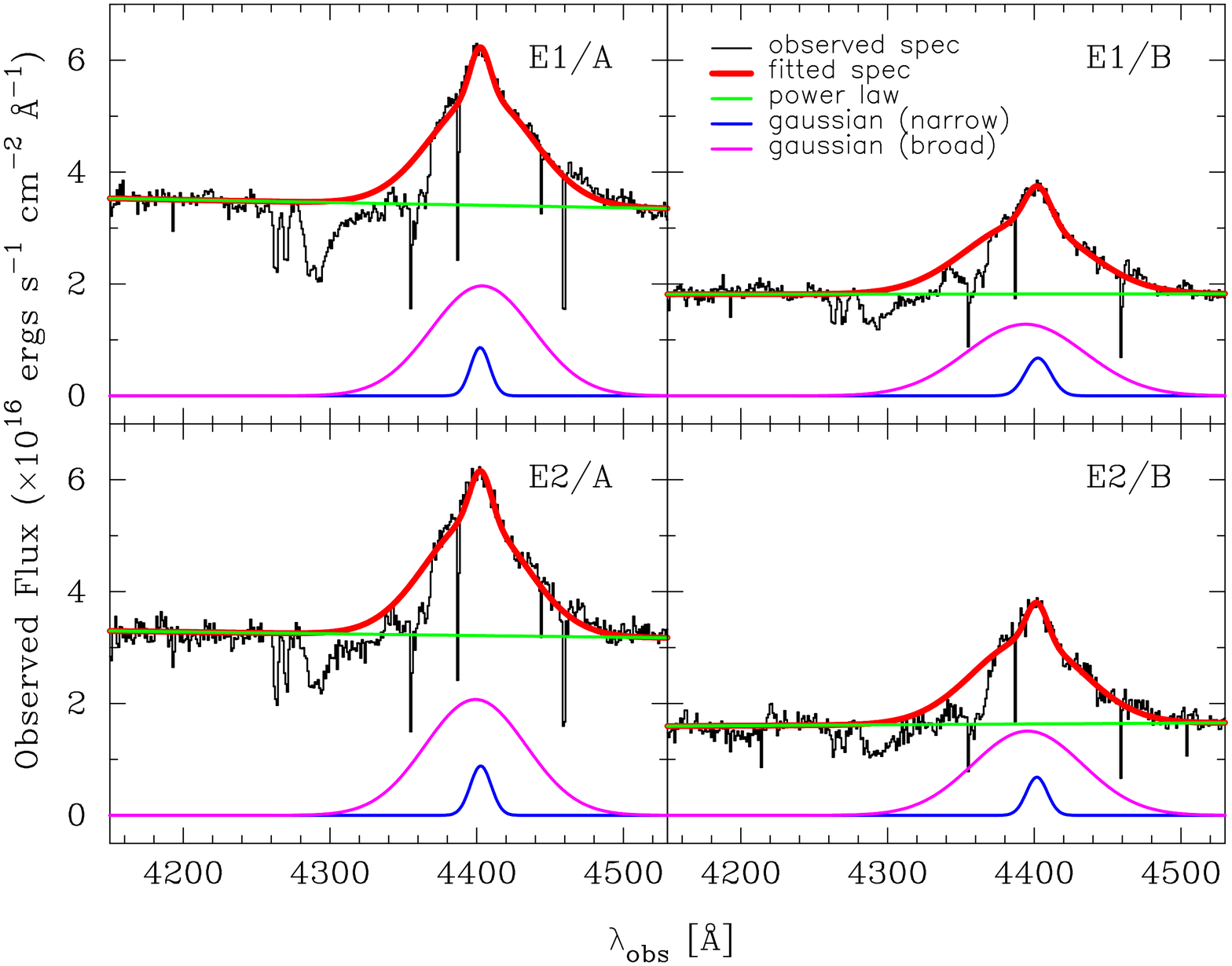}
\end{center}
\caption{Flux calibrated spectra around the \ion{C}{4} emission lines
  of images~A and B of the quasar SDSS~J1001+5027 in epoch~E1 (upper
  left and right) and epoch~E2 (lower left and right), with
  multi-component fits using two Gaussian components (i.e., broad and
  narrow emission lines) and power-law (i.e., continuum level).  The
  green line shows the continuum fit, the magenta and blue lines show
  the two Gaussian components and the thick red line shows the sum of
  all of the above.\label{fig:belfit}}
\bigskip
\end{figure}

We show normalized spectra over the full wavelength range of our
observations for images~A and B in Figure~\ref{fig:spec}.  These
spectra were binned every 0.5~\AA\ for display purposes, and the
1$\sigma$ errors are also shown.  Because the blue wing of the
\ion{C}{4} emission line is severely absorbed at $\lambda$
$\sim$4300~\AA, we cannot fit the continuum in that region directly.
Instead, we model the continuum and emission line profile using a
power law and two Gaussian profiles for the broad and narrow
\ion{C}{4} emission components (Figure~\ref{fig:belfit}), which is a
standard practice.  We do not consider the \ion{He}{2}~$\lambda$1640
emission line that is located about 250\AA\ redward of the \ion{C}{4}
emission line, because its contribution to the blue wing of the
\ion{C}{4} emission line is negligible.  The best-fitting models are
very similar for the four spectra; FWHMs of broad and narrow emission
components are 5890$\pm$390~\kms\ and 1180$\pm$125~\kms, respectively,
which are close to the typical values of optically bright quasars at
$z$$\sim$2, $\sim$4900~\kms\ and $\sim$1400~\kms\ \citep[e.g.,][and
  references therein]{browills94,mar96,sul17}.

% =================================================================
% Figure 2
% =================================================================
% =================================================================

In addition to the broad absorption feature above, we also detect
narrower absorption features whose confidence level is greater than
5$\sigma$ in the normalized spectra of images~A and B in both
epochs~E1 and E2, using the line detection code {\sc search}, written
by Chris Churchill.  We first identify doublet lines including
\ion{C}{4} and \ion{Mg}{2} in the regions between the bluest edge of
the observed spectrum ($\lambda$ $\sim$ 3820\AA) and the corresponding
emission lines.  In total, we identified one \ion{Mg}{2} and three
\ion{C}{4} systems in the image~A spectrum, and two \ion{Mg}{2} and
three \ion{C}{4} systems in the image~B spectrum in both epochs.  At
the redshift of these systems, we also detected 11 single metal
absorption lines (\ion{Si}{2}~$\lambda$1527,
\ion{Fe}{2}~$\lambda$1608, \ion{Al}{2}~$\lambda$1671,
\ion{Al}{3}~$\lambda$1855, \ion{Al}{3}~$\lambda$1863,
\ion{Fe}{2}~$\lambda$2344, \ion{Fe}{2}~$\lambda$2374,
\ion{Fe}{2}~$\lambda$2383, \ion{Fe}{2}~$\lambda$2587,
\ion{Fe}{2}~$\lambda$2600, and \ion{Mg}{1}~$\lambda$2853), as
summarized in Table~\ref{tab:abslines}.  The \ion{Mg}{2} system at the
redshift of the lensing galaxy (\zabs\ $\sim$ 0.415) is detected only
in the image~B spectrum, which is probably why the image~B spectrum is
redder compared to the image~A spectrum, as noted in \citet{ogu05}.

% =================================================================
% Table 2
% =================================================================
\begin{deluxetable*}{ccccccccrrr}
\tablecaption{Narrow Absorption Lines in epoch E1 \label{tab:abslines}}
\tablewidth{0pt}
\tablehead{
\multicolumn{1}{c}{}                   &
\multicolumn{3}{c}{Image~A}            &
\multicolumn{1}{c}{}                   &
\multicolumn{3}{c}{Image~B}            &
\multicolumn{3}{c}{}                   \\
\cline{2-4} 
\cline{6-8}
\colhead{Ion}                          &
\colhead{\zabs}                        &
\colhead{EW$_{\rm obs}$$^a$}             &
\colhead{ID}                           &
\colhead{}                             &
\colhead{\zabs}                        &
\colhead{EW$_{\rm obs}$$^a$}             &
\colhead{ID}                           &
\colhead{$D_{\perp}$$^b$}               &
\colhead{$\Delta v$$^c$}               &
\colhead{$|\Delta$EW$|^d$}             \\
\colhead{}                             &
\colhead{}                             &
\colhead{(\AA)}                        &
\colhead{}                             &
\colhead{}                             &
\colhead{}                             &
\colhead{(\AA)}                        &
\colhead{}                             &
\colhead{(pkpc)}                       &
\colhead{(\kms)}                       &
\colhead{(\AA)}                       
}
\startdata
\ion{Mg}{2}$\lambda$2796     & ...    & ...             &    &  & 0.4147 & 2.205$\pm$0.150 &  1 & 15.7 &      ... &             ... \\ 
\ion{Mg}{2}$\lambda$2803     & ...    & ...             &    &  &        & 1.663$\pm$0.124 &  3 &      &          &             ... \\ 
\hline                                                                                              
\ion{Mg}{2}$\lambda$2796     & ...    & ...             &    &  & 0.4166 & 0.352$\pm$0.043 &  2 & 15.7 &      ... &             ... \\ 
\ion{Mg}{2}$\lambda$2803     & ...    & ...             &    &  &        & 0.214$\pm$0.042 &  4 &      &          &             ... \\ 
\hline                                                                                              
\ion{Mg}{2}$\lambda$2796     & 0.8716 & 1.949$\pm$0.058 & 19 &  & 0.8714 & 1.656$\pm$0.105 & 23 &  7.0 &     32.1 & 0.293$\pm$0.120 \\ 
\ion{Mg}{2}$\lambda$2803     &        & 1.638$\pm$0.055 & 20 &  &        & 1.279$\pm$0.090 & 24 &      &          & 0.359$\pm$0.105 \\ 
\ion{Fe}{2}$\lambda$2344     &        & 0.806$\pm$0.046 & 12 &  &        & 0.529$\pm$0.050 & 16 &      &          & 0.277$\pm$0.068 \\ 
\ion{Fe}{2}$\lambda$2374     &        & 0.283$\pm$0.025 & 13 &  &        & 0.213$\pm$0.050 & 17 &      &          & 0.070$\pm$0.056 \\ 
\ion{Fe}{2}$\lambda$2383     &        & 1.271$\pm$0.051 & 14 &  &        & 1.029$\pm$0.096 & 18 &      &          & 0.242$\pm$0.109 \\ 
\ion{Fe}{2}$\lambda$2587     &        & 0.557$\pm$0.030 & 16 &  &        & 0.414$\pm$0.057 & 20 &      &          & 0.143$\pm$0.064 \\ 
\ion{Fe}{2}$\lambda$2600     &        & 1.134$\pm$0.048 & 18 &  &        & 0.648$\pm$0.080 & 22 &      &          & 0.486$\pm$0.093 \\ 
\ion{Mg}{1}$\lambda$2853     &        & 0.457$\pm$0.042 & 21 &  &        & 0.322$\pm$0.036 & 25 &      &          & 0.135$\pm$0.055 \\ 
\hline
\ion{C}{4}$\lambda$1548      & 1.6067 & 3.076$\pm$0.171 &  2 &  & 1.6067 & 3.213$\pm$0.264 &  6 &  1.0 &      0.0 & 0.137$\pm$0.315 \\ 
\ion{C}{4}$\lambda$1551      &        & 2.307$\pm$0.160 &  3 &  &        & 2.190$\pm$0.232 &  7 &      &          & 0.117$\pm$0.282 \\ 
\ion{Si}{2}$\lambda$1527     &        & 0.664$\pm$0.036 &  1 &  &        & 0.628$\pm$0.045 &  5 &      &          & 0.036$\pm$0.058 \\ 
\ion{Fe}{2}$\lambda$1608     &        & 0.183$\pm$0.022 &  8 &  &        & 0.235$\pm$0.057 & 12 &      &          & 0.052$\pm$0.061 \\ 
\ion{Al}{2}$\lambda$1671$^e$ &        & 0.677$\pm$0.031 & 11 &  &        & 0.670$\pm$0.046 & 15 &      &          & 0.007$\pm$0.055 \\ 
\ion{Al}{3}$\lambda$1855     &        & 0.525$\pm$0.038 & 15 &  &        & 0.567$\pm$0.069 & 19 &      &          & 0.042$\pm$0.079 \\ 
\ion{Al}{3}$\lambda$1863     &        & 0.445$\pm$0.034 & 17 &  &        & 0.522$\pm$0.062 & 21 &      &          & 0.077$\pm$0.071 \\ 
\hline                                                                                              
\ion{C}{4}$\lambda$1548      & 1.6505 & 0.149$\pm$0.034 &  4 &  & 1.6530 & 0.248$\pm$0.042 &  8 &  0.8 & $-$282.8 & 0.099$\pm$0.054 \\ 
\ion{C}{4}$\lambda$1551      &        & 0.060$\pm$0.016 &  5 &  &        & 0.149$\pm$0.032 &  9 &      &          & 0.089$\pm$0.036 \\ 
\hline                                                                                              
\ion{C}{4}$\lambda$1548      & 1.6766 & 0.129$\pm$0.028 &  6 &  & 1.6766 & 0.386$\pm$0.053 & 10 &  0.7 &      0.0 & 0.257$\pm$0.060 \\ 
\ion{C}{4}$\lambda$1551      &        & 0.070$\pm$0.023 &  7 &  &        & 0.219$\pm$0.050 & 11 &      &          & 0.149$\pm$0.055 \\ 
\hline                                                                                              
\ion{C}{4}$\lambda$1548$^f$  & 1.7535 & 1.208$\pm$0.149 &  9 &  & 1.7536 & 1.069$\pm$0.205 & 13 &  0.3 &  $-$10.9 & 0.139$\pm$0.253 \\ 
\ion{C}{4}$\lambda$1551$^f$  &        & 1.109$\pm$0.192 & 10 &  &        & 0.993$\pm$0.234 & 14 &      &          & 0.116$\pm$0.303 
\enddata
\tablenotetext{a}{Observed-frame equivalent width.}
\tablenotetext{b}{Physical distance between two sightlines in the
  transverse direction.}
\tablenotetext{c}{Velocity Offset between NALs detected in images~A
  and B (\vej($z_B$) $-$ \vej($z_A$)).}
\tablenotetext{d}{Equivalent width difference between NALs detected in
  images~A and B.}
\tablenotetext{e}{This line is blended with \ion{C}{4} BAL.}
\tablenotetext{f}{This line is probably a part of \ion{C}{4} BAL.}
\end{deluxetable*}
% =================================================================

\subsection{Broad Absorption Line}
The broad P-Cygni profile of the \ion{C}{4} emission line was known
from medium resolution spectra ($R$ $\sim$ 2000) of the quasar
\citep{ogu05}.  This broad absorption feature with FWHM of
$\sim$7000~\kms\ appears fairly smooth in our high-resolution
spectrum, which means that the feature is indeed intrinsically broad
and smooth and cannot be attributed to intervening absorbers whose
typical line width is much smaller (i.e., up to several hundreds of
\kms).

Because this feature satisfies the definition of a broad absorption
line (BALs), with more than 10\%\ of the flux absorbed over at least a
2000~\kms\ range continuously \citep{wey91}, we refer to it as a BAL,
hereafter.  Corresponding BAL features in lines other than \ion{C}{4}
are not detected at the same redshift in our spectra although the
wavelength region corresponding to the \ion{Si}{4} BAL is severely
affected by bad CCD columns\footnote[17]{There is no obvious broad
  absorption features detected in intermediate resolution spectra ($R$
  $\sim$ 2000) at which \ion{Si}{4} BAL is expected to be located
  \citep{mor17}.}.  In a UV spectrum of the quasar taken with HST/COS,
\citet{mor17} also detected \ion{O}{6} (and possibly \ion{P}{5}) BALs.
The offset velocity from the quasar emission redshift along the line
of sight (i.e., the apparent ejection velocity) of the \ion{C}{4} BAL
is \vej\ $\sim$ 2500--9500~\kms\ (positive values correspond to
blueshifts).

\begin{figure}[h]
 \begin{center}
  \includegraphics[width=8cm,angle=0]{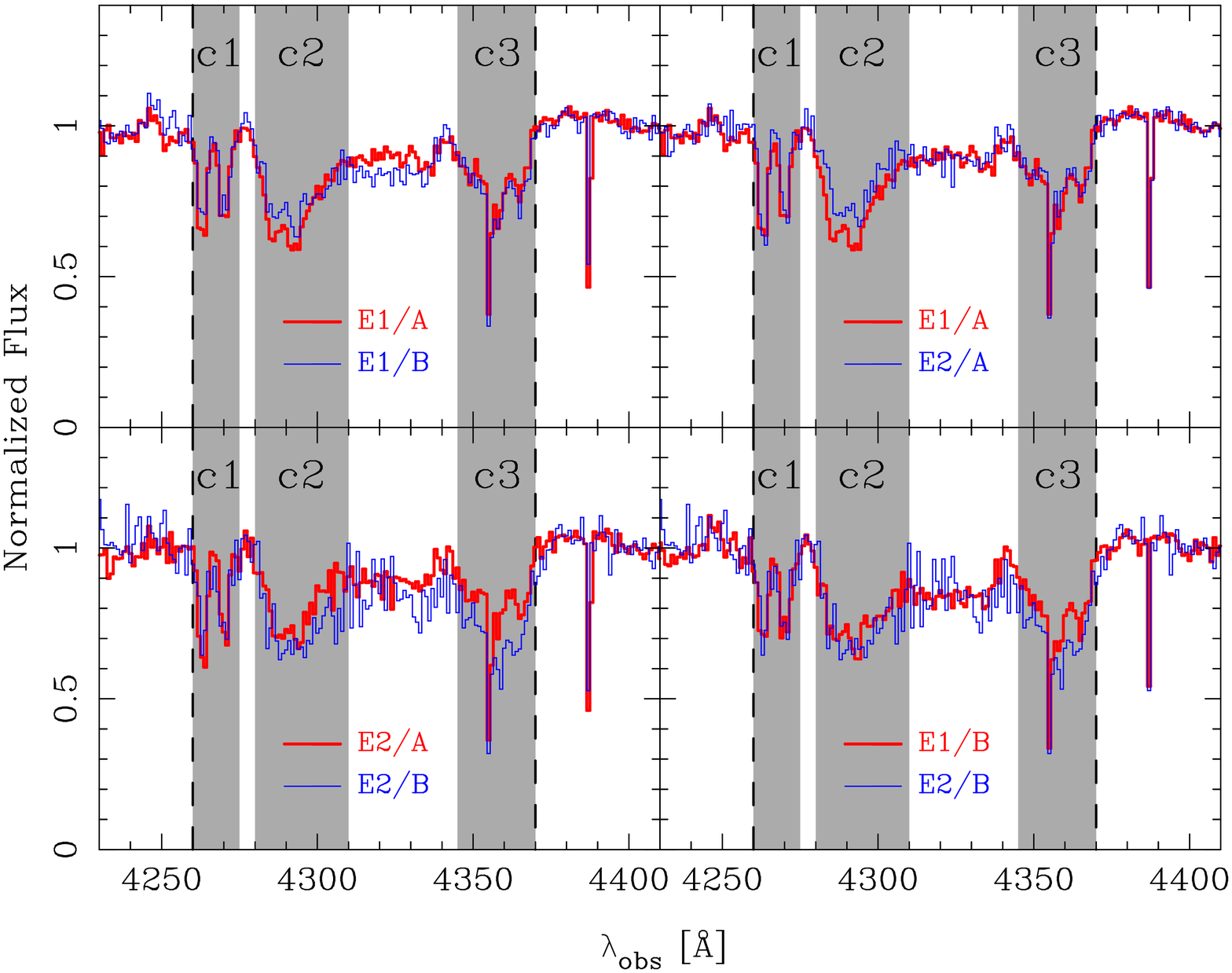}
 \end{center}
\caption{Comparison of \ion{C}{4} BAL profiles in the normalized
  spectra between image~A and B in epoch E1 (upper left) and E2 (lower
  left) and between image~A (upper right) and image~B (lower right) in
  epochs E1 and E2. Three shaded areas denote the components c1, c2,
  and c3.\label{fig:balcomp}}
\end{figure}

Examining the \ion{C}{4} BAL in more detail, we note that the feature
consists of three separate troughs; one narrow and two broader
components.  We will call these BAL components c1, c2, and c3, as
shown in Figure~\ref{fig:balcomp}.  Only component c1 has the blue and
red members of the doublet unblended from each other. A sharp feature
in component c3 at $\lambda$ $\sim$4355~\AA\ is not part of the BAL
feature but is instead an \ion{Al}{2}~$\lambda$1671 line from an
intervening absorber at \zabs\ = 1.6067.

% =================================================================
% Figure 3
% =================================================================
\vspace{1cm}
% =================================================================

%%% Time Variability %%%
We also monitored the variability of the absorption profile and
strength of the \ion{C}{4} BAL using spectra in two epochs whose time
separation is 358~days in the observed frame (i.e., $\sim$126 days in
the quasar rest frame).  Because the two cylinders of sight to the two
lensed images overlap completely at the distance of the absorber (as
we demonstrate later in \S 5.1), we effectively monitored the same
absorber at four epochs using the observations of the two lensed
images in two epochs; the relative time intervals of the four epochs
are $\Delta t_{\rm obs}$ = 0, 119, 358, and 476~days.

We evaluated the equivalent width (EW) of the \ion{C}{4} BAL with its
1$\sigma$ error, $\sigma$(EW), in our four spectra, to search for
variability.  Because our measurements of EW suffer not only
photon-noise ($\sigma_{\rm phot}$) but also uncertainties in the
placement of the continuum level ($\sigma_{\rm cont}$), we combine
these two errors in quadrature to obtain the final 1$\sigma$ error.
Following \citet{mis14a}, the latter error is calculated empirically
as
\begin{equation}
  \sigma_{\rm cont}(EW) = \frac{(\lambda_{\rm max}-\lambda_{\rm
      min})}{2\;(S/N)},
\label{eqn:sigcont}
\end{equation}
where $\lambda_{\rm max}-\lambda_{\rm min}$ is the wavelength width of
the \ion{C}{4} BAL \citep[see the discussion in][]{mis14a}.

\begin{figure}[h]
 \begin{center}
  \includegraphics[width=8cm,angle=0]{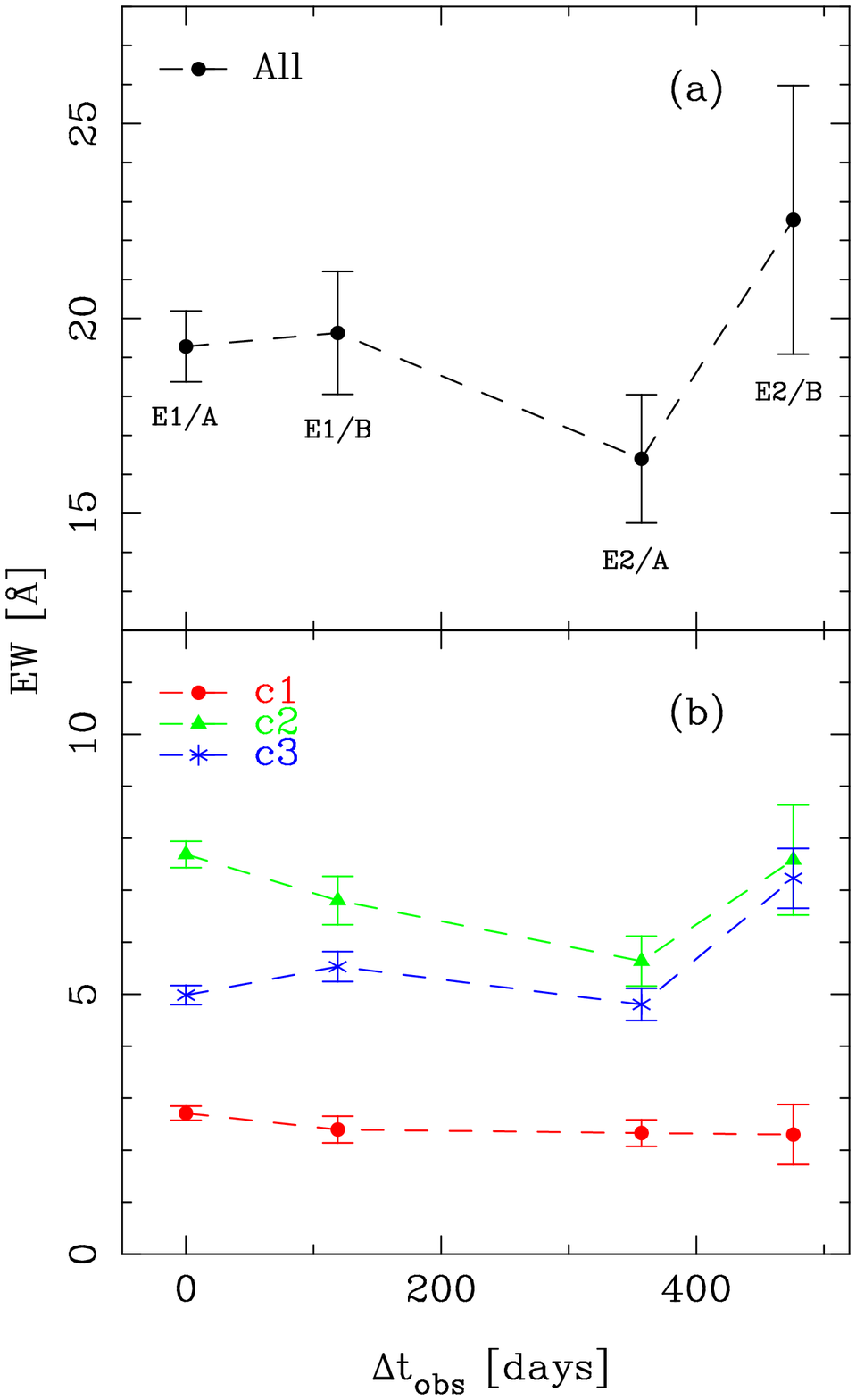}
 \end{center}
\caption{Monitoring result for EWs of whole range of the \ion{C}{4}
  BAL (top) and each component (bottom). The horizontal axis is the
  time delay from the first epoch in the observed frame (in days).
  The vertical axis is the observed EWs (with 1$\sigma$
  errors).\label{fig:ewvar}}
\end{figure}

We do not find clear variability in the total equivalent width of the
\ion{C}{4} BAL between observing epochs, although there is a small
difference between images~A and B in epoch~E2 as shown in
Figure~\ref{fig:ewvar}a.  However, once we separate the BAL feature
into three components and measure their EW separately, we note that
the broader components c2 and c3 show clear variability between epochs
(see Figures~\ref{fig:balcomp} and \ref{fig:ewvar}b).  It appears that
the EW of the component c2 first decreased in 358~days and then
increased in 119~days, while that of component c3 is almost stable in
the first 358~days but suddenly increased in 119~days.  Thus, the
variability pattern is not synchronized.  We will discuss a plausible
scenario to explain this behavior in the following section.

% =================================================================
% Figure 4
% =================================================================
\vspace{1.0cm}
% =================================================================

\begin{figure}[h]
 \begin{center}
  \includegraphics[width=8cm,angle=0]{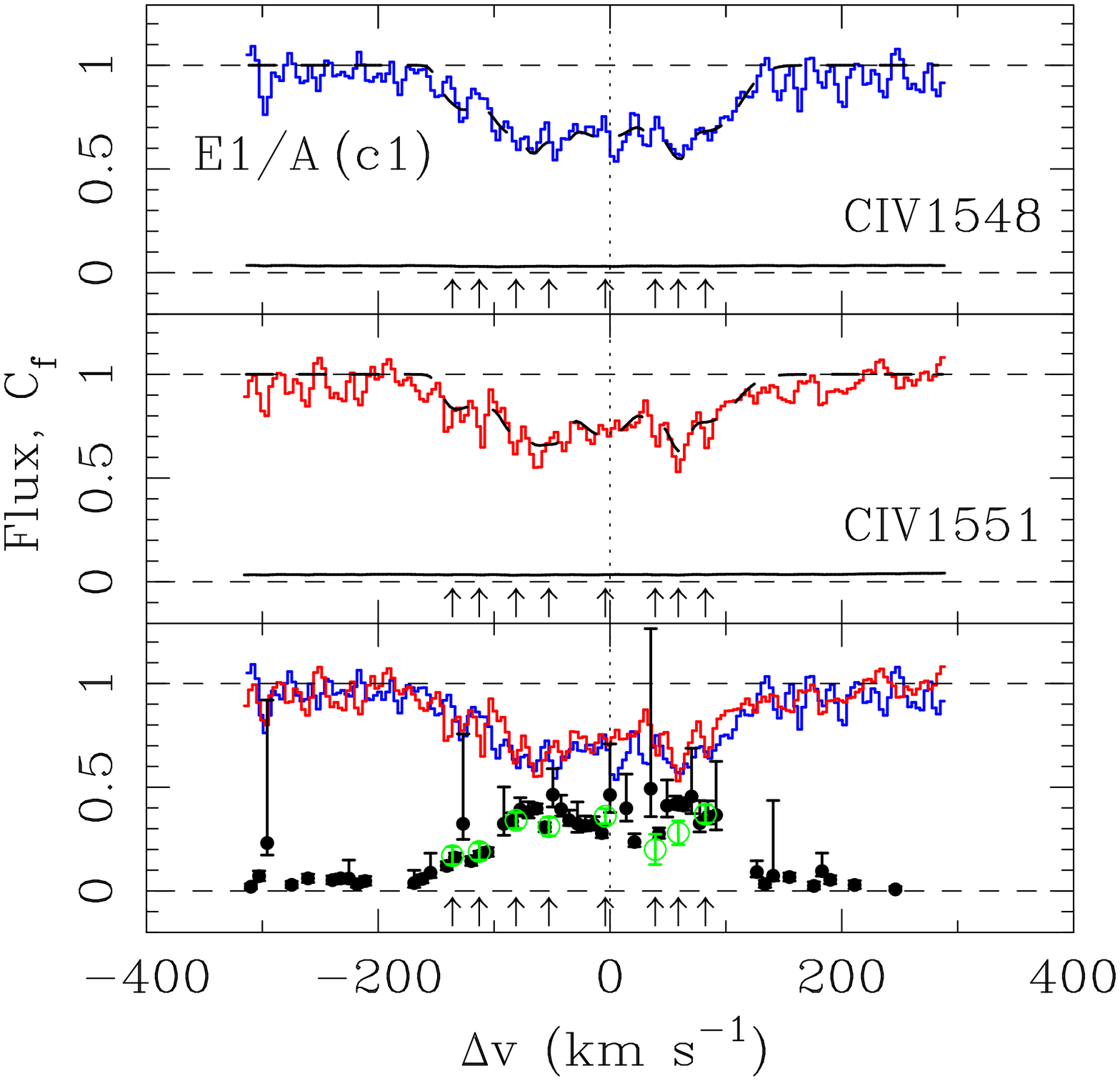}
 \end{center}
 \caption{Results of partial coverage analysis applied to the
   component c1 of the \ion{C}{4} BAL detected in image~A of
   SDSS~J1001+5027 in epoch E1. The horizontal axis denotes the
   relative velocity from the flux-weighted center of the system
   ($\Delta v$) while the vertical axis is the normalized flux. The
   first two panels show the profiles of the blue and red members of a
   doublet (blue and red histograms) with the model profile produced
   by {\sc minfit} superposed (dashed line). The positions of the
   absorption components are marked with upward arrows in the bottom
   of each panel.  The bottom panel shows the covering factors with
   their 1$\sigma$ errors, measured for each narrow component by {\sc
     minfit} (green circles) or for each bin (black
   dots).\label{fig:minfit}}
\end{figure}

%%% Covering Factor Analysis for comp c1 %%%
Of three components of the \ion{C}{4} BAL, only component c1 has the
blue and red members of the doublet unblended, which enables us to
apply partial coverage analysis. We examine whether the optical depth
ratio of the doublet lines to see if it deviate from the value of 2:1
prescribed by atomic physics. Attributing any such deviation to
dilution of absorption troughs by unocculted light from the background
source allows us to find the fraction of the projected area of the
source that is occulted by the absorber
\citep[e.g.,][]{wam95,bar97,ham97a,gan99}.  We can evaluate a covering
factor (\cf), a fraction of background light occulted by the absorber,
by using the equation,
\begin{equation}
  \cf(\lambda) =
  \frac{\left[1-R_r(\lambda)\right]^{2}}{1+R_b(\lambda)-2R_r(\lambda)},
  \label{eqn:cf}
\end{equation}
where $R_{b}$ and $R_{r}$ are the continuum normalized intensities of
the stronger (bluer) and weaker (redder) members of the doublet
\citep[see, for example][]{bar97}.  We evaluate \cf\ values for each
      {\it bin} (not for each absorption component) as done in
      \citet{gan99}. We can also use the line fitting software package
            {\sc minfit} \citep{chu97,chu03} to fit absorption
            profiles with four parameters; absorption redshift
            (\zabs), column density ($\log N$ in \cmm), Doppler
            parameter ($b$ in \kms), and covering factor (\cf).  If
            {\sc minfit} gives unphysical covering factors for some
            components, we rerun the code assuming \cf\ = 1 for those
            components following \citet{mis05}.  As an example, the
            fitting result for the image~A spectrum in epoch~E1 by
            both methods is shown in Figure~\ref{fig:minfit}. The fit
            parameters for all four spectra are summarized in
            Table~\ref{tab:minfit}.  Although the number of absorption
            components needed to fit the absorption profile depends on
            the S/N of each spectrum (from eight components for
            image~A in epoch~E1 to only one component for image~B in
            epoch~E1), the covering factors evaluated for each bin and
            for each absorption component are always consistent with
            each other in all spectra.  This result gives us
            confidence that component c1 shows partial coverage with
            \cf\ $\sim$ 0.3, which is smaller than the value evaluated
            for \ion{O}{6} BAL detected in the UV spectrum
            (\cf\ $\sim$ 0.67; \citealt{mor17}).

% =================================================================
% Table 3
% =================================================================
\begin{deluxetable*}{cccccc}
\tablecaption{Line Parameters of \ion{C}{4} BAL \label{tab:minfit}}
\tablewidth{0pt}
\tablehead{
\colhead{epoch/image$^a$} & 
\colhead{\zabs$^b$} &
\colhead{\vej$^c$} &
\colhead{$\log N$} &
\colhead{$b$$^d$} &
\colhead{\cf$^e$} \\
\colhead{} &
\colhead{} &
\colhead{(\kms)} &
\colhead{(\cmm)} &
\colhead{(\kms)} &
\colhead{} 
}
\startdata
E1/A & 1.7524 &  9507 & 14.12$\pm$0.31 &  7.8$\pm$2.0 & 0.17$^{+0.04}_{-0.04}$ \\
     & 1.7526 &  9484 & 14.02$\pm$0.17 & 11.2$\pm$2.6 & 0.19$^{+0.04}_{-0.04}$ \\
     & 1.7529 &  9452 & 14.12$\pm$0.09 & 13.2$\pm$1.5 & 0.34$^{+0.04}_{-0.04}$ \\
     & 1.7531 &  9424 & 14.45$\pm$0.19 & 10.8$\pm$1.4 & 0.31$^{+0.05}_{-0.05}$ \\
     & 1.7536 &  9375 & 14.35$\pm$0.06 & 27.6$\pm$3.2 & 0.36$^{+0.04}_{-0.04}$ \\
     & 1.7540 &  9332 & 13.95$\pm$0.30 &  7.3$\pm$2.6 & 0.20$^{+0.08}_{-0.08}$ \\
     & 1.7541 &  9312 & 14.20$\pm$0.43 &  5.7$\pm$1.8 & 0.28$^{+0.06}_{-0.06}$ \\
     & 1.7544 &  9289 & 14.12$\pm$0.10 & 29.2$\pm$3.3 & 0.37$^{+0.05}_{-0.04}$ \\
\cline{1-6}
E1/B & 1.7536 &  9373 & 15.53$\pm$0.15 & 72.3$\pm$6.0 & 0.29$^{+0.05}_{-0.05}$ \\
\cline{1-6}
E2/A & 1.7531 &  9431 & 14.60$\pm$0.09 & 37.6$\pm$3.5 & 0.31$^{+0.05}_{-0.05}$ \\
     & 1.7534 &  9395 & 16.63$\pm$3.23 & 10.6$\pm$6.0 & 0.14$^{+0.06}_{-0.06}$ \\
     & 1.7542 &  9306 & 14.57$\pm$0.05 & 45.2$\pm$2.4 & 0.42$^{+0.05}_{-0.05}$ \\
\cline{1-6}
E2/B & 1.7536 &  9375 & 13.90$\pm$0.21 & 25.0$\pm$3.2 & 0.63$^{+0.25}_{-0.20}$ \\
     & 1.7542 &  9311 & 16.26$\pm$2.72 &  9.2$\pm$4.8 & 0.36$^{+0.09}_{-0.09}$ 
\enddata
\tablenotetext{a}{Observed epoch (E1/E2) and lens image (A/B).}
\tablenotetext{b}{Redshift of flux weighted line center.}
\tablenotetext{c}{Ejection velocity from quasar emission redshift.}
\tablenotetext{d}{Doppler parameter.}  
\tablenotetext{e}{Covering factor.}  
\end{deluxetable*}
% =================================================================

% =================================================================
% Figure 5
% =================================================================
\vspace{1.0cm}
% =================================================================

\subsection{Narrow Absorption Lines}
While BALs are very likely physically associated with quasars, NALs
are more difficult to classify as intrinsic or intervening.  There are
several reliable criteria to distinguish intrinsic NALs from
intervening NALs including time variability, line locking, and partial
coverage \citep[e.g.,][]{bar97,ham97a}. Among these, the first two are
applicable to our spectra, while we cannot use partial coverage
analysis because our NALs are detected in regions of our spectra where
the S/N is low.\footnote[18]{We can apply the analysis only for
  component c1 of the \ion{C}{4} BAL above because of its simple and
  deep absorption profile located in a higher S/N region.}  None of
the NALs show time variability (with one possible exception below) nor
line-locking, while both of these effects are seen in several NALs in
the spectra of the large-separation lensed quasar SDSS~J1029+2623
\citep{mis16}.

In addition to these three methods of identifying intrinsic NALs,
\citet{mis16} introduced a new method by exploiting multi-sightline
observations.  If NALs with identical line profiles are detected at
the same redshift (i.e., at a common apparent ejection velocity from
the quasar) along two sightlines with a large physical distance in the
transverse direction, the corresponding absorber is very likely to be
intrinsic to the quasar, because both cylinders of sight to the two
lensed images can go through the same gas parcels only if these
parcels are located in the vicinity of the flux source.  For example,
if an absorber is located at a distance of $\sim$1~pc from the source,
the physical separation between the two sightlines
($\sim$10$^{-5}$~pc) is smaller than the typical size of a clump in
the outflow, $d$ $\sim$10$^{-4}$~pc, as estimated in \citet{ham13}.
However, this method is effective only if ($i$) it is applied to {\it
  metal} absorption lines such as \ion{C}{4} and \ion{Mg}{2}
NALs\footnote[19]{\ion{H}{1} absorbers in the IGM (i.e., \lya\ forest)
  have a much larger projected size ($\sim$1~Mpc) than the physical
  distance between two lines of sight to SDSS~J1001+5027
  ($\sim$1--10~kpc; see Table~\ref{tab:abslines}).}, ($ii$) the NALs
to be compared are detected in same quality spectra (i.e., same
spectral resolution, S/N ratio, and sampling), and ($iii$) the
physical distance between two sightlines in the transverse direction
at the apparent redshift of the absorber is comparable to or larger
than the typical size of the metal absorbers (i.e., $\sim$0.05--5~kpc;
\citealt{ste16})\footnote[20]{Intervening galaxies and their
  circum-galactic media (CGM) are the main intervening metal
  absorption line systems. Although the typical extent of the CGM from
  galaxies is $\sim$200--300~kpc \citep[e.g.,][]{che01,chu13,tur14},
  their covering factor is not unity \citep[i.e., they have an
    inhomogeneous internal structure; e.g.,][]{kac12,bor14,koy17}. The
  typical size of {\it each} clump is estimated to be $\sim$4~kpc or
  $\sim$50~pc for \ion{Si}{4} and \ion{Mg}{2} absorbers, respectively
  \citep{ste16}. We can use the \ion{Si}{4} absorber size as a proxy
  for the \ion{C}{4} absorber size because of its similar ionization
  potential.  Even if two cylinders of sight to the two lensed images
  pass through a single cloud, the corresponding absorption-line
  profiles in their spectra are not identical because their total
  column densities are unlikely to be same (e.g., one goes through the
  middle of the cloud, while the other does not.).}  Our spectra of
SDSS~J1001+5027 satisfy all of these requirements.  Although some NAL
pairs are detected at almost the same redshift along the two
sightlines, their line profiles are clearly different from each other.
Since none of the NAL profiles match between sightlines A and B toward
SDSS~J1001+5027, we cannot identify any of the 6 NAL systems detected
in the spectra as intrinsic ones based only on this method.

% =================================================================
% Figure 6
% =================================================================
\begin{figure*}
 \begin{center}
  \includegraphics[width=12cm,angle=270]{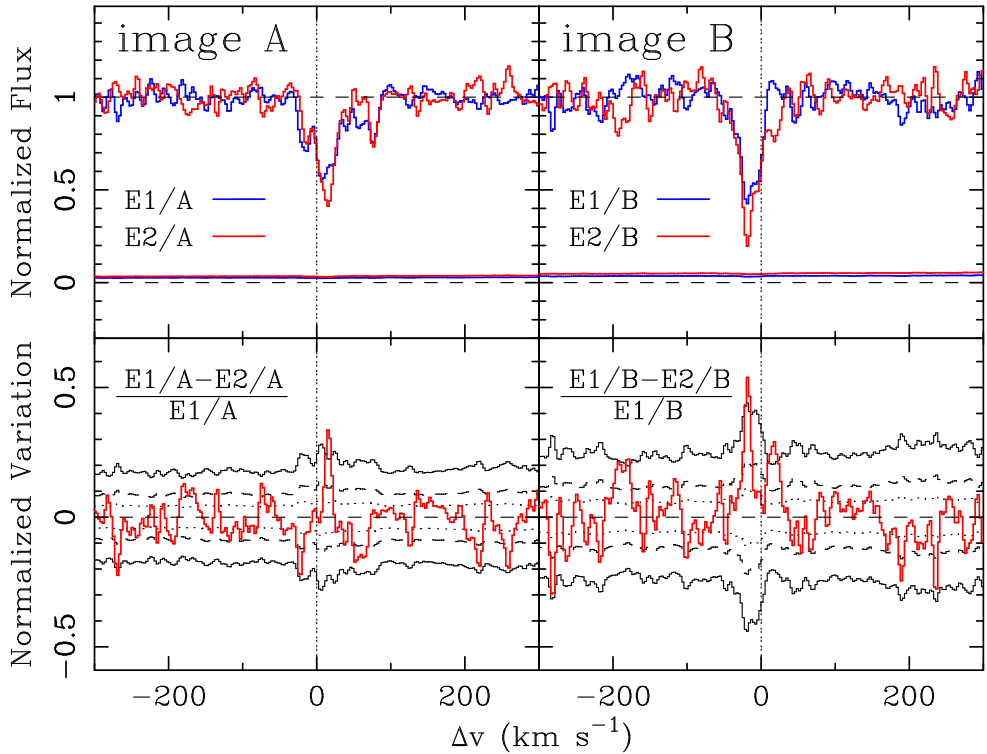}
 \end{center}
 \caption{Comparison of normalized spectra around the \ion{Mg}{1} NAL
   detected at \zabs\ = 0.8716 in spectra of the images~A (upper left)
   and B (upper right) in epochs E1 (blue) and E2 (red).  The
   histograms above the zero flux line are 1$\sigma$ flux errors.
   Normalized flux variations between the spectra in two epochs as a
   function of velocity ((E1$-$E2)/E1; solid histogram) are shown for
   images~A (lower left) and B (lower right) with 1, 2, and 4$\sigma$
   flux uncertainties (dotted, dashed, and solid curves).  The
   horizontal axis is the offset velocity from \zabs\ =
   0.8716.\label{fig:nalvar}}
\end{figure*}
% =================================================================

\begin{figure*}
 \begin{center}
  \includegraphics[width=18cm,angle=270]{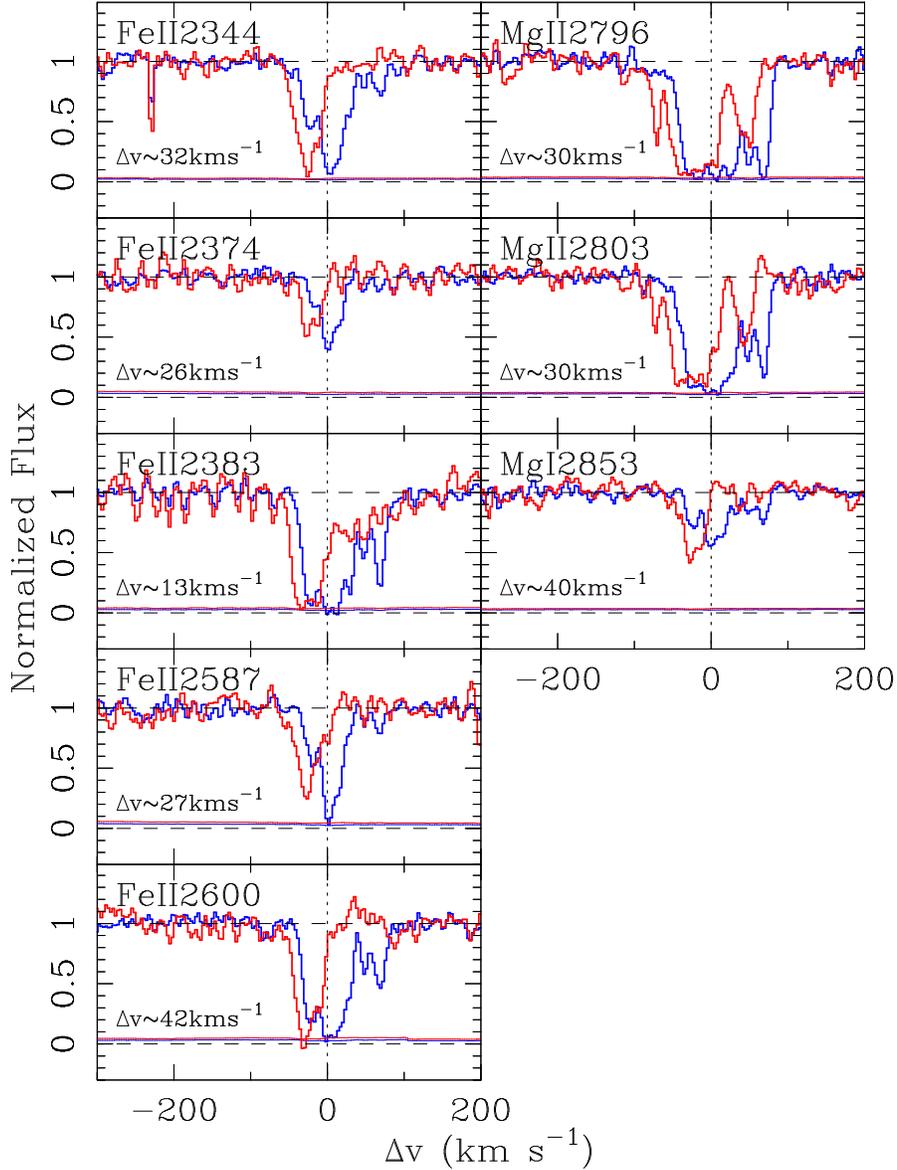}
 \end{center}
 \caption{Comparison of normalized spectra around detected NALs of the
   \ion{Mg}{2} system at \zabs\ $\sim$ 0.8716 in spectra of the
   images~A (blue) and B (red) in epoch E1. The horizontal axis is the
   offset velocity from the flux-weighted line center of \ion{Mg}{2}
   in the image~A (\zabs\ = 0.8716), and the vertical axis is the
   normalized flux.  The histograms above the zero flux line are
   1$\sigma$ flux errors.  The numbers at the lower left corner of
   each panel denote the offset velocity of the NALs in image~A from
   that in image~B.\label{fig:nalcomp}}
\end{figure*}

%%% Gas motion in a foreground galaxy %%%
Only the \ion{Mg}{1} NAL of the \ion{Mg}{2} absorption system at
\zabs\ = 0.8716 shows a hint of time variability with differences at
the $\sim$4$\sigma$ significance level near the minimum of the line
profile. The profile appears to become narrower in $\Delta t_{\rm
  obs}$ = 358~days ($\Delta t_{\rm rest}$ = 191~days at \zabs\ =
0.8716) between epochs~E1 and E2 (Figure~\ref{fig:nalvar}).  However,
the detection of variability is marginal because there are several
noise spikes in the spectra near the \ion{Mg}{1} NAL with a similar
significance level of variability.  We need higher quality spectra to
confirm the \ion{Mg}{1} variability.  It is also remarkable that all
NALs in the system at \zabs\ = 0.8716 including the \ion{Mg}{1} NAL
show a velocity offset in the sense that NALs in the image~A are
redshifted from those in the image~B between two sightlines with a
projected separation of $\sim$7~kpc (Figure~\ref{fig:nalcomp}). We
computed the flux-weighted line center as the first moment of the line
profile for each NAL in the spectra of both images, and found that the
line centers in both spectra have substantial offset velocities with
$\Delta v$ = 30$\pm$8~\kms.  We interpret both the {\it marginal}
\ion{Mg}{1} variability and the velocity offset as the result of
motion of gas within or around a foreground galaxy at \zabs\ = 0.8716.
For example, the NAL system could originate in gas clouds that are
tidally stripped by a galaxy merger or accreting onto or outflowing
from a foreground galaxy.  This would be one of only a few cases in
which NALs of low-ionization species show any time variability
\citep[e.g., ][and references therein]{hac13}.

% =================================================================
% Figure 7
% =================================================================
% =================================================================

\section{Discussion}
\subsection{Broad Absorption Line}
The \ion{C}{4} BAL shows variability in profile and strength between
the two observing epochs along both sightlines.  There are at least
three possible origins for the variability; gas motion across our
sightline \citep[e.g.,][]{ham08,gib08,viv12,viv14,viv16,kro17}, a
change of ionization state of the absorber
\citep[e.g.,][]{ham11,ara12,mis07b,tre13,fil13,hor16}, and a
microlensing event \citep[e.g.,][]{gre06}.  Among these, the last one
is less likely because the microlensing effect of BELs in
SDSS~J1001+5027 is almost negligible \citep{gue13}.  The observed
variability trends can place further constraints on the possible
scenarios.  The time variability is seen in only portions of the
\ion{C}{4} BAL and its variability pattern depends on the absorber's
ejection velocity from the source (Figure~\ref{fig:balcomp}).  A
change of incident ionizing flux, which would affect all components in
concert, is clearly inconsistent with the observed behavior. In
principle, the ionization state of the absorber could change because
of changes in its density due to expansion or compression but such
changes are very slow in view of the local sound speed. We estimate
the sound speed to be $c_s\sim10\, T_4^{1/2}\;\kms$, where $T_4$ is
the temperature scaled by $10^4\;$K, which is a typical value of the
temperature of a photoionized gas. Using the covering factor evaluated
above and assuming conservatively that only the continuum source is
covered by the absorber we find that the absorber size is $a >
\cf\,R_{\rm cont} \sim 7\times 10^{-4}\;$pc. Hence the characteristic
time for changes in the density due to internal motions in the
absorber is $\ge 70\;$years, which is two orders of magnitude longer
than the observed variability time scale. The mismatch in time scales
remains significant even if we attribute different troughs of the
profile to different portions of the absorber; such portions must be
substantial fractions of the entire absorber as the profile comprises
only three troughs.  Therefore, as the most plausible and natural
explanation we take the bulk gas motion scenario that is typically
adopted in the case that variability occurs in just a portions of BAL
profiles like SDSS~J1001+5027 \citep[e.g.,][]{gib08,ham08,cap13}.

In our preferred scenario for the variability the absorber is moving
either in front of the continuum source or in front of the BELR at the
local dynamical speed.  The latter is implausible because it would
imply a transverse (i.e., rotational) velocity greater than the speed
of light, namely $v_{\rm rot} \geq R_{\rm BELR}/\Delta t_{\rm rest} >
c$. Thus, the continuum source is the only plausible background source
in this context, and the inferred rotational velocity is $v_{\rm rot}
\geq R_{\rm cont}/\Delta t_{\rm rest}$. Taking $\Delta t_{\rm
  rest}=42\;$days, the interval over which the most substantial EW
variation is observed (see Figure~\ref{fig:ewvar}), we find that
$v_{\rm rot} \geq 18,200\;$\kms.  Assuming Keplerian motion, the
absorber's distance is then $r\leq 0.06$~pc. Based on intermediate
resolution spectra ($R$ $\sim$ 2000), \citet{mor17} also discovered
variability in the \ion{C}{4} BAL strength at the $4\sigma$ confidence
level on a time scale of $\sim$0.29~yrs in the quasar rest frame,
which gives a limit on the absorber's distance ($r < 0.4\;$pc), which
is consistent with our estimate. According to this estimate, the
absorber is in the inner parts of the BELR. Moreover, if we assume
that the absorber's observed velocity offset is indeed its outflow
velocity, then the absorber is bound to the black hole.

The inferred absorber distance is much smaller than the boundary
distance of the continuum source ($r_{\rm b}$ $\sim$ 325~pc).  Thus,
the cylinders of sight to the two lensed images overlap almost
completely at the absorber's distance. In other words, they represent
a common sightline, and we have indeed monitored the BAL profile at
four epochs.

A consequence of this scenario is that an absorber so close to the
ionizing source should be highly ionized unless its density is also
quite high.  For example, the volume density of the BAL absorber is
required to be $n_{\rm e}$ $\sim$ 10$^{13}$~\cmmm, if we assume a
typical value of the ionization parameter for the C$^{3+}$ ion ($\log
U$ $\sim$ $-$1.7) \citep{ham95}.  This is much larger than the
expected value (e.g., $n_{\rm e}$ $\sim$ 10$^8$~\cmmm\ in a mini-BAL
absorber) \citep{ham13}.  Nonetheless, there are known cases of dense
absorbers located very close to the ionizing continuum source, e.g.,
Arp~102B \citep{era03} and FIRST~J104459.6+365605
\citep{dek01}. Another possibility is that there exists some shielding
material between the continuum source and the absorber. The shielding
material could be the warm absorber frequently detected in X-ray
spectroscopy \citep[e.g.,][]{gal02,kro07}.

% =================================================================
% Table 4
% =================================================================
\begin{deluxetable*}{ccccc}
\tablecaption{Comparison of Two Lensed Quasars\label{tab:comp}}
\tablewidth{0pt}
\tablehead{
\colhead{} & 
\colhead{SDSS~J1029+2623} &
\colhead{Ref.$^a$} &
\colhead{SDSS~J1001+5027} &
\colhead{Ref.$^a$}
}
\startdata
\zem$^b$                             & 2.197               & 1      & 1.841                & 10    \\
\zl$^c$                              & 0.58                & 2      & 0.415                & 11    \\
$\theta$ (arcsec) $^d$               & 22.5                & 1      & 2.86                 & 10    \\
$\theta^{\prime}$ (arcsec) $^e$        & 14.6                & 3      & 1.37                 & 12    \\
$\mu_{\rm A}$ $^f$                     & 10.4                & 4     & 1.99                 & 10,12  \\
$\Delta t_{\rm obs}$ (days) $^g$       & 744                  & 5     & 119                  & 13     \\
$\log L_{\rm bol}$ (erg s$^{-1}$) $^h$ & 45.87                & 6,7   & 46.93                & 7,12   \\
$L_{\rm bol}/L_{\rm Edd}$ $^i$           & 0.11                 & 6     & 0.15                 & 12     \\
$\log M_{\rm BH}/M_{\odot}$ $^j$        & 8.72                 & 6,7   & 9.66                 & 7,12   \\
$R_{\rm BELR}$ (pc) $^k$               & 0.09                 & 6,7,8 & 0.37                 & 7,8,12 \\
$R_{\rm cont}$ (pc) $^l$               & 2.54$\times$10$^{-4}$ & 6,7,9 & 2.16$\times$10$^{-3}$ & 7,9,12 \\
\hline
\vej (\kms) $^m$                     & 0 -- 1600            & 6     & 2500 -- 9500         & 12     \\
FWHM (\kms) $^n$                     & 1600                 & 6     & 7000                 & 12     \\
Absorption Profile $^o$              & NAL                  & 6     & BAL                  & 12     \\
Covering factor                      & \cf\ $<$ 1           & 6     & \cf\ $<$ 1           & 12     \\
Line locking                         & Yes                  & 6     & No                   & 12     \\
$r_b$(BELR) (kpc) $^p$               & 1.2                  & 3      & 55.6                & 12     \\
$r_b$(cont) (pc) $^q$                & 3.5                  & 3      & 325                 & 12     
\enddata
\tablenotetext{a}{References. (1) \citet{ina06}, (2) \citet{ogu08},
  (3) \citet{mis16}, (4) \citet{ogu13}, (5) \citet{foh13}, (6)
  \citet{mis13}, (7) \citet{she11}, (8) \citet{mcl04}, (9)
  \citet{mis05}, (10) \citet{ogu05}, (11) \citet{ina12}, (12) this
  paper, (13) \citet{rat13}}
\tablenotetext{b}{Quasar emission redshift.}
\tablenotetext{c}{Redshift of lensing galaxy.}
\tablenotetext{d}{Observed separation angle.}
\tablenotetext{e}{Separation angle seen from the source.}
\tablenotetext{f}{Magnification factor of the brightest image.}
\tablenotetext{g}{Time delay between images~A and B.}
\tablenotetext{h}{Bolometric luminosity.}
\tablenotetext{i}{Eddington Ratio.}
\tablenotetext{j}{Black hole mass.}
\tablenotetext{k}{Size of \ion{C}{4} broad emission line region.}
\tablenotetext{l}{Size of the continuum source.}
\tablenotetext{m}{Ejection velocity of outflowing wind.}
\tablenotetext{n}{FWHM of associated absorption system.}
\tablenotetext{o}{Category of absorption profile (BAL or NAL).}
\tablenotetext{p}{Boundary distance considering BELR.}
\tablenotetext{q}{Boundary distance considering only the continuum source.}
\end{deluxetable*}
% =================================================================

\subsection{Narrow Absorption Lines}
There are three non-associated\ \ion{C}{4} NAL systems along both of
the two sightlines toward SDSS~J1001+5027 (\ion{C}{4} pairs,
hereafter). These are systems with apparent ejection velocities
greater than 5,000 \kms\ from the quasar emission redshift
\citep[e.g.,][]{wey79}. In \citet{mis16}, we also detected 11
non-associated \ion{C}{4} pairs toward the large-separation lensed
quasar SDSS~J1029+2623 \citep{mis16}.  Among these 14 \ion{C}{4} pairs
some could still be physically associated with the quasars even if
their ejection velocity is greater than 5,000~\kms. Indeed,
\citet{mis07a} estimated that about 20\%\ of non-associated \ion{C}{4}
NALs are intrinsic to quasars based on a covering factor analysis.
However, we found that no \ion{C}{4} pairs have common absorption
profiles, which means all \ion{C}{4} systems (even if some of them are
physically associated with the quasar) are located further from the
source than the boundary distance, $r_{\rm b}$.  Nonetheless, if some
systems are intrinsic to the quasar, these results would be consistent
with the geometrical model of the outflowing wind in
\citet{mis14b,mis16}, in which broader absorption systems like BALs
are located at $r \leq r_{\rm b}$, while intrinsic NALs are at $r \gg
r_{\rm b}$.  As for NALs with a large ejection velocity, only the
continuum source can be the background source because their position
in the spectrum is far away from the corresponding BELs.  We detected
neither time variability (with one possible exception discussed in the
next paragraph) nor line-locking of any NAL systems as signs of
intrinsic absorption lines in the non-associated lines in our spectra
of SDSS~J1001+5027.  The latter suggests our sightline is not parallel
to the outflow wind that is radiatively accelerated with resonance
line absorption even if some of them are intrinsic to the quasar. This
is in contrast to the variable, intrinsic NALs found in quasar
SDSS~J1029+2623 \citep{mis13,mis16}.

The only absorption line that shows a hint of time variability in both
strength and profile is the \ion{Mg}{1} NAL of the \ion{Mg}{2}
absorption system at \zabs\ = 0.8716 (Figure~\ref{fig:nalvar}). This
system is less likely to be intrinsic to the quasar because its offset
velocity from the quasar (\vej\ $\sim$ 118,000~\kms) is quite large.
Until now, there are only a few cases of variable intervening
absorption lines with large offset velocities of \vej\ $>$
66,000~\kms\ \citep[][and references therein]{hac13}.

Assuming variability of the \ion{Mg}{1} NAL is due to gas motion
across the sightlines, we can place an upper limit on the cloud's size
as $d \leq v_{\rm trans} \, \Delta t_{\rm rest} \sim 22~(v_{\rm
  trans}/200~\kms)\;$AU, where we normalized the transverse velocity
by the typical rotational velocity of intervening galaxies.  This
system could be an analog of low-ionized ($\log U \leq -7$) metal
absorbers in a small sized ($\sim$ 10--100~AU), cold ($<$ 100~K), and
high volume density ($n_{\rm H}$ $\sim$ 10$^3$--10$^6$~\cmmm) pocket
of gas with a large molecular content which have been detected in the
Milky Way \citep[e.g.,][]{dia89,fai98,ric03} as well as outside our
Galaxy \citep[e.g.,][]{kan01,jon10,hac13,ber17}.

Another mechanism for the possible variability of the \ion{Mg}{1} NAL
is a change of ionization state.  In this case, we can place
constraints on the electron density as $n_e$ $\geq$ 5,800 \cmmm,
assuming that the decrease in the ionizing flux causes recombination
from Mg$^{+}$ to Mg$^{0}$ \citep[e.g.,][]{ham97a,nar04}.  However, a
change in ionization state is less likely because it rarely occurs in
just a few years in foreground galaxies \citep{nar04}.

\subsection{Multi-Sightline Observation for Lensed Quasars}
We performed high-resolution spectroscopic observations for a small
separation lensed quasar twice in this paper and for a
large-separation lensed quasar SDSS~J1029+2623 three times in
\citet{mis13,mis14b,mis16}.  The parameters we evaluated for the
quasars based on the above observations are summarized in
Table~\ref{tab:comp}.  As seen in Figure~10 of \citet{mis16} and
Figure~\ref{fig:balcomp} of this paper, the absorption profiles in the
two lensed images are similar, which means that in both cases there is
a significant overlap of the two cylinders of sight to the two lensed
image. Thus, the corresponding absorbers are probably located at a
distance smaller than the boundary distance \citep[see Figure~13
  of][]{mis16}.

\subsubsection{Intrinsic NALs in SDSS~J1029+2623}
As for the NAL clustering at \zabs\ $\sim$ \zem\ in the spectra of
SDSS~J1029+2623 \citep[we called it a Proximity Absorption Line, or
  PAL, system in][]{mis16}, we discovered that all the troughs in the
PAL varied in concert, which suggests the observed variability was not
due to gas motion.  Assuming the variability is due to a change of
ionization, \citet{mis16} placed an upper limit on its distance from
the source of $r < 620$~pc by interpreting the variability time scale
to be the recombination time.  This distance is consistent with those
evaluated for other intrinsic NAL systems in quasars that are not
lensed, a few hundred parsecs to several kiloparsec
\citep[e.g.,][]{ham01,nar04,ara13,bor13}. At such a large distance,
the BELR can be a background flux source that contributes along with
the continuum source.  Indeed, the residual flux at the bottom of the
PAL is almost zero on the \ion{C}{4} emission line in the spectra.
The boundary distance, assuming that both the continuum source and the
BELR are background sources, is $r_{\rm b} \sim 1.2$~kpc
\citep{mis16}.  Thus, the NAL absorbers are viewed through distinct
sightlines in large-separation lensed quasars with $\theta \geq
10^{\prime\prime}$ because their distance from the flux source is
large enough (i.e., comparable to the boundary distance).

Small-separation lensed quasars with $\theta\sim 1^{\prime\prime}$ may
still provide views through distinct sightlines of NALs with very
large offset velocities that are located far from the corresponding
BELs in quasar spectra (i.e., only the continuum source is the
background source.).

\subsubsection{BAL in SDSS~J1001+5027}
In contrast, BAL absorbers are usually located at a very small radial
distance of order a parsec or much less \citep[e.g.,][]{cap11,cap13}.
Assuming that the variability we have observed in the \ion{C}{4} BAL
of SDSS~J1001+5027 is due to a gas motion, we placed an upper limit on
the absorber's distance as $r \leq 0.06$--0.5~pc.  At such a small
distance, the BAL absorbers could co-exist with the BELR \citep[][see
  also \citealt{era03}]{kro17}.  Even if only the continuum source is
the background source, the corresponding boundary distance is much
larger than the absorber's distance (see Table~\ref{tab:comp}).  Thus,
a multi-sightline observation is almost impossible for BAL absorbers
in the vicinity of the BELR.  However, large-separation lensed quasars
may allow us to perform multi-sightline observation even for BAL
absorbers if the absorbers are located far from the BELR and if the
BAL troughs do not overlap with the BEL profiles.

\section{Summary}
Using the Subaru telescope, we performed high-resolution spectroscopy
of the two images of the small-separation lensed quasar
SDSS~J1001+5027 ($\theta = 2.^{\!\!\prime\prime}86$) twice in a time
interval of about one year to see whether the broad \ion{C}{4}
absorption lines are variable or not.  We discuss the constraints on
the radial distance and physical conditions of the absorber and
compare them to those obtained for the large-separation lensed quasar
SDSS~J1029+2623 ($\theta = 22.^{\!\!\prime\prime}5$). We also search
for variability and compare profiles between the sightlines for NAL
absorbers seen in our spectra.  Our main findings are:

\begin{itemize}
\item We discovered variability in the \ion{C}{4} BALs in both lensed
  images.  The observed variations are probably caused by gas motion
  around the background source (i.e., the continuum source) because
  the absorption components in the BAL profile do not vary in
  concert. We infer a rotational velocity of the absorber of $v_{\rm
    rot}$ $\geq$ 18,000~\kms, and a radial distance from the center of
  $r$ $\leq$ 0.06~pc assuming Keplerian motion.
\item Because BAL absorbers are believed to be located very close to
  the flux source (at a distance of order parsecs or less, much
  smaller than the boundary distance), the sightlines of the images of
  small-separation lensed quasars, such as SDSS~J1001+5027, are not
  distinct. Nonetheless, the sightlines to NAL absorbers that are
  several kiloparsec away from the quasar are distinct \citep[as in
    SDSS~J1029+2623][]{mis13,mis14b,mis16}.
\item In addition to the \ion{C}{4} BAL system, we detected 6 NAL
  systems at \zabs\ = 0.4147--1.7536, of which only the \ion{Mg}{2}
  system at \zabs\ = 0.8716 shows possible time variability in its
  \ion{Mg}{1} profile and a $\sim$30~\kms\ velocity shear between the
  sightlines whose physical separation is $\sim$ 7~kpc. This is
  probably due to motion of the absorbing gas located in an
  intervening galaxy.
\end{itemize}

For further investigations of the outflow's geometrical and physical
conditions through multi-sightline observations, we should mainly
target intrinsic NAL absorbers with a large offset velocity from
quasar redshift (i.e., their distance from the flux source is large)
rather than BAL systems in the vicinity of the source.  As for BALs,
we can use lensed quasar spectra taken at a single epoch to examine
their variability, exploiting the time delay between the lensed
images.

\acknowledgments We would like to thank the anonymous referee for
extensive comments, that helped improve the paper.  We also would like
to thank Ken Ohsuga and Mariko Nomura for their valuable comments.  We
also would like to thank Christopher Churchill for providing us with
the {\sc minfit} and {\sc search} software packages.  The research was
supported by the Japan Society for the Promotion of Science through
Grant-in-Aid for Scientific Research 15K05020 and partially supported
by a MEXT Grant-in-Aid for Scientific Research on Innovative Areas
(No. 15H05894).  JCC and ME were supported by National Science
Foundation grant AST-1312686.

\end{document}